\documentclass{aa}
\input epsf
\usepackage{graphicx}
\usepackage[mathcal]{eucal}
\usepackage{amssymb}
\usepackage{amsmath}
\usepackage{natbib}
\usepackage[T1]{fontenc}
\usepackage{color}

\def\gtrsim{\mathrel{\hbox{\rlap{\hbox{\lower4pt\hbox{$\sim$}}}\hbox{$>$}}}}
\def\ltsim{\mathrel{\hbox{\rlap{\hbox{\lower4pt\hbox{$\sim$}}}\hbox{$<$}}}}

\def\kms{\hbox{{\rm km}\,{\rm s}$^{-1}$}}
\def\kms {km\,s$^{-1}$}

\begin{document}

\title{Discovery of a complex linearly polarized spectrum of Betelgeuse dominated by depolarization of the continuum 
\thanks{Based on observations obtained at the T\'elescope Bernard Lyot (TBL) at Observatoire du Pic du Midi, CNRS/INSU and Universit\'e de Toulouse, France.}}

\author{M. Auri\`ere\inst{1,2}, A. L\'opez Ariste\inst{1,3}, P. Mathias\inst{1,2}, A. L\`ebre\inst{4}, E. Josselin\inst{4}, M. Montarg\`es \inst{5,6}, P. Petit\inst{1,3},\\ 
A. Chiavassa\inst{7}, F. Paletou\inst{1,3}, N. Fabas\inst{8}, R. Konstantinova-Antova\inst{9}, J.-F. Donati\inst{1,3}, J.H. Grunhut\inst{10}, G.A. Wade\inst{11}, F. Herpin\inst{12,13}, P. Kervella\inst{6,14}, G. Perrin\inst{6}, B. Tessore\inst{4}}
\institute{Universit\'e de Toulouse, UPS-OMP, Institut de Recherche en Astrophysique et Plan\'etologie, Toulouse, France
\email{michel.auriere@irap.omp.eu}
\and
CNRS, UMR5277, Institut de Recherche en Astrophysique et Plan\'etologie, 57 Avenue d'Azereix,65000 Tarbes, France
\and
CNRS, UMR5277, Institut de Recherche en Astrophysique et Plan\'etologie, 14 Avenue Edouard Belin, 31400 Toulouse, France
\and
LUPM - UMR5299 - CNRS and Universit\'e Montpellier  - Place E. Bataillon, 34090 Montpellier, France
\and
Institut de Radioastronomie Millim\'etrique, 300 rue de la Piscine, 38406, Saint Martin d'H\`eres, France 
\and
LESIA, Observatoire de Paris, PSL Research University, CNRS UMR 8109, Sorbonne Universit\'es, UPMC, Universit\'e Paris Diderot, Sorbonne Paris Cit\'e,
5 place Jules Janssen, F-92195 Meudon, France 
\and
Laboratoire Lagrange, Universit\'e C\^ote d'Azur, Observatoire de la C\^ote d'Azur, CNRS,   Blvd de L'Observatoire CS 34229,06004 Nice Cedex 4, France
\and
Kiepenheuer-Institut for Sonnenphysik, 79 104 Freiburg, Germany
\and
Institute of Astronomy and NAO, Bulgarian Academy of Sciences, 72 Tsarigradsko shose, 1784 Sofia, Bulgaria
\and
European Southern Observatories, Karl-Schwarzschild-Str. 2, D-85748, Garching, Germany
\and
Department of Physics, Royal Military College of Canada,
  PO Box 17000, Station `Forces', Kingston, Ontario, Canada K7K 4B4
\and
Universit\'e de Bordeaux, LAB, UMR 5804, F-33270, Floirac, France
\and
CNRS, LAB, UMR 5804, F-33270, Floirac, France
\and
Unidad Mixta Internacional Franco-Chilena de Astronom\'ia (UMI 3386), CNRS/INSU, France \& Departamento de
Astronom\'ia, Universidad de Chile, Camino El Observatorio 1515, Las Condes, Santiago, Chile.
}

\date{Received ??; accepted ??}

\abstract
{\object{Betelgeuse} is an M supergiant that harbors spots and giant granules at its surface and presents linear polarization of its continuum.}
{We have previously discovered linear polarization signatures associated with individual lines in the spectra of cool and evolved stars. Here, we 
investigate whether a similar linearly polarized spectrum exists for \object{Betelgeuse}.}
{We used the spectropolarimeter Narval, combining multiple polarimetric sequences to obtain high signal-to-noise ratio spectra of individual lines, as well as the least-squares deconvolution 
(LSD) approach, to investigate the presence of an averaged linearly polarized profile for the photospheric lines.}
{We have discovered the existence of a linearly polarized spectrum for \object{Betelgeuse}, detecting a rather strong signal (at a few times 10$^{-4}$ of the 
continuum intensity level), both in individual lines and in the LSD profiles.
Studying its properties and the signal observed for the resonant \ion{Na}{i}\,D lines, we conclude that we are mainly observing 
depolarization of the continuum by the absorption lines. 
The linear polarization of the Betelgeuse continuum is due to the anisotropy of the radiation field induced by brightness 
spots at the surface and Rayleigh scattering in the atmosphere. 
We have developed a geometrical model to interpret the observed polarization, from which we infer the presence of two brightness spots and their positions on the surface of
\object{Betelgeuse}. We show that applying the model to each velocity bin along the Stokes $Q$ and $U$ profiles allows the derivation of a map of the bright spots.
We use the Narval linear polarization observations of \object{Betelgeuse} obtained over a period of 1.4 years to study the evolution of the spots and of the atmosphere.}
{Our study of the linearly polarized spectrum of \object{Betelgeuse} provides a novel method for studying the evolution of brightness spots at its surface and
complements quasi-simultaneous observations obtained with PIONIER at the VLTI.}

   \authorrunning {M. Auri\`ere et al.}
   \titlerunning {The linearly polarized spectrum of Betelgeuse}

\maketitle

\section{Introduction}

The study of the linearly polarized spectrum of the Sun enabled the discovery of a very rich variety of structures which opened up `a 
new window for diagnostics of the Sun' \citep{s97}. 
This spectrum is composed of structures resulting partly from intrinsic polarization of the lines and also from the depolarization 
of the continuum  by the absorption lines. 
The existence of this `second solar spectrum' is due to the existence of anisotropy of the radiation field, induced by limb darkening. 
In the stellar case, it is considered to be very difficult to observe the second solar spectrum, since in general distant stars present a symmetric aspect 
\citep[e.g.][]{lag11}. 
However, in the case of cool and evolved low- and intermediate-mass stars, linear polarization has already been reported in individual lines 
of variable Mira stars: in the Balmer hydrogen lines \citep{mc78,flg11,f11} and in the resonance line of \ion{Ca}{i} at 422.6\,nm \citep{bac86}.
In addition \citet{laf14} discovered as yet unexplained nonzero Stokes $Q$ and $U$ signals when averaging a mean profile about 10\,000 atomic lines of the Mira star \object{$\chi$\,Cyg}
observed at a luminosity maximum, using the least squares deconvolution (LSD) approach \citep{dsc97}. 
Similar LSD signals as well as intrinsic polarization in individual lines such as \ion{Na}{i} D2 and \ion{Sr}{i} at 460.7\,nm were discovered in other Miras and RV\,Tauri stars \citep{laf15}. 
It therefore appears that anisotropy of the radiation field may also exist in very evolved cool stars, and that a rich linearly polarized spectrum does exist in these pulsating variable stars.

We report here the subsequent study of the linearly polarized spectrum of \object{Betelgeuse} (\object{$\alpha$\,Ori}, \object{HD\,39801}) 
which is a nearby M2\,Iab supergiant, and which may be considered as a high-mass counterpart to the evolved AGB stars (as are the Miras). 
Being one of the stars with the largest apparent diameters, it has naturally deserved direct imaging and interferometric studies of its surface, 
which was found to deviate from circular symmetry and to present variable behavior \citep{wdh97,hpl09,ker15}. 
A surface magnetic field has also been detected which appears to vary on a timescale of weeks/months 
\citep[Mathias et al. in preparation]{adk10,pak13,bpa13}.
In this paper, we demonstrate that Stokes $QU$ observations reveal a rich linearly polarized spectrum suitable for analysis using both individual and averaged spectral line profiles. 

In Section\,2 we present the Narval spectropolarimetric observations, in Section\,3 we describe the detection of the linear polarization signal and we show 
that the detected signal is of stellar origin. 
In Section\,4 we infer the origin of the polarization together with some of its properties.
Finally, we present a geometric model in Section\,5,  study the variations of the linear polarization with time in Section\,6, and give our conclusions in Section\,7.

\section{Observations with Narval}

\begin{table*}
\caption{Log of observations of Betelgeuse and polarimetric measurements (for details, see Section\,2 and Section\,5). }          
\label{tab1}   
\centering                         
\begin{tabular}{l c l c c c c c c c c}     
\hline\hline               
Date             & Stokes  & ${P_L}_1$  & ${\theta}_1$ & ${\chi}_1$ & ${\mu}_1$  & ${P_L}_2$ & ${\theta}_2$    & ${\chi}_2$ & ${\mu}_2$ \\
                 &         & 10$^{-4}$  &  $^\circ$     &  $^\circ$    & $^\circ$    &  10$^{-4}$ & $^\circ$        & $^\circ$    & $^\circ$ \\
\hline                        
27 November 2013 $ \in$ Set\,1 & $2Q 2U$     & 6.8      & 23.2        & 113.2       &  73.0     &  3.4      & 122.0          & 212.0      & 84.0\\
11 December 2013 $ \in$ Set\,1 & $2Q 2U$     & 8.4      & 22.0        & 112.0       &  73.0     &  4.3      & 120.1          & 210.1      & 81.9\\
20 December 2013 $ \in$ Set\,1 & $2Q 2U$     & 8.3      & 22.7        & 112.7       &  73.0     &  4.7      & 121.3          & 211.3      & 81.9\\
09 January  2014 $ \in$ Set\,1 & $2Q 2U$     & 6.0      & 24.9        & 114.9       &  70.8     &  4.9      & 124.5          & 214.5      & 84.0\\  
08 April    2014 & $1Q 1U$     & 4.7      & 26.7        & 116.7       &  66.4     &  4.2      & 103.7          & 193.7      & 81.9 \\
12-13 September 2014$ \equiv$ Set\,2 & $8Q 8U$     & 5.8      & 39.4        & 129.4       &  73.0     &  3.5      & 119.2          & 209.2      & 84.0\\
16 October 2014  & $8Q 8U$     & 4.5      & 35.2        & 125.2       &  79.4     &  3.4      & 103.3          & 193.3      & 79.4\\
23 October 2014  & $8Q 8U$     & 3.6      & 37.0        & 127.0       &  81.4     &  3.6      & 103.3            & 193.3      & 81.4\\
20 November 2014 & $8Q 8U$     & 4.9      & 45.0        & 135.0       &  79.4     &  2.8      & 112.5          & 202.5      & 77.7\\
18 December 2014 & $8Q 8U$     & 5.1      & 51.9        & 141.9       &  77.2     &  2.9      & 125.3          & 215.3      & 79.8\\
03 March 2015    & $8Q 8U$     & 2.7      & 68.7        & 158.7       &  77.2     &  2.9      & 146.9          & 236.9      & 77.7 \\
13 April 2015    & $8Q 8U$     & 5.1      & 90.4        & 180.4       &  75.1     & 2.2       & 161.8           & 251.8      & 75.6 \\
\hline
\hline                            
\end{tabular}
\tablefoot{Columns present the date, the number of Stokes $QU$ spectra obtained, then for the two spots (spot1 and spot2 modeled in Section 5) the observed maximum of linear polarization $P_L $, polarization angle $\theta$, position angle $\chi$, and projection angle $\mu$.}
\end{table*} 

Observations of \object{Betelgeuse} were obtained at the 2m T\'elescope Bernard Lyot (TBL) using the Narval spectropolarimeter, which is a twin of 
the ESPaDOnS instrument at the Canada-France-Hawaii Telescope (CFHT) \citep{d03,dcl06}. 
The observations that include linear polarization span from November 2013 to April 2015. 
A standard polarization observation consists of a series of four sub-exposures between which the half-wave retarders 
(Fresnel rhombs) are rotated to exchange the paths of the orthogonally-polarized beams within the whole 
instrument (and therefore the positions of the two spectra on the Charge-Couple Device (CCD), thereby reducing spurious polarization signatures. 
To avoid saturation of the CCD, we used 3\,s exposure times for each sub-exposure. 
The extraction of the spectra, including wavelength calibration, correction to the heliocentric frame and
continuum normalisation, was performed using the Libre-ESpRIT package \citep{dsc97}, installed both at CFHT and
TBL.
The extracted spectra are in ASCII format, and consist of the normalized Stokes $I$ ($I/I_{\rm c}$) spectrum and 
Stokes $Q, U$ ($Q/I_{\rm c}$, $U/I_{\rm c}$) parameters as a function of wavelength, together with their associated 
uncertainties (where $I_{\rm c}$ represents the continuum intensity). In the Libre-ESpRIT reduction the continuum polarization level is automatically removed. In the case of \object{Betelgeuse} for which a polarized continuum does exist (see Section 4.1), the level of polarization of the continuum cannot be recovered accurately from our spectra since Narval is not designed for this purpose. 
Also included in the output are `diagnostic null' spectra $N$, which are in principle featureless, and therefore serve 
to diagnose the presence of spurious contributions to the polarized spectra. 
Each single spectrum used in this work has a peak signal-to-noise ratio (S/N) in Stokes $I$ per 1.8\,\kms\ spectral 
bin between 1700 and 2100. 
Details of the observation and reduction procedures are provided by \citet{dsc97} and \citet{adk10,awk09}.
Table\,\ref{tab1} reports the log of observations. 

To obtain a high precision diagnosis of the spectral line polarization, the least-squares
deconvolution approach \citep[LSD,][]{dsc97} was applied to each reduced Stokes $I$, $Q$ and $U$ spectrum. 
We used a solar abundance line mask calculated from data provided by the Vienna Atomic Line Database \citep[VALD,][]{kpr99} for an effective 
temperature of 3750\,K, surface gravity $\log g =0.0$, and a microturbulence of 4.0\,\kms, consistent with the physical parameters 
of \object{Betelgeuse} \citep{jp07,lbh84}. 
The complete mask contains about 15\,000 atomic lines with a central depth greater than 40\% of the continuum (which are below 60\% of the continuum). 
Specific elements with transitions expected to trace a shocked region, a chromosphere or a circumstellar medium  
(such as \ion{H}, \ion{Ca}{i}, \ion{Na}{i}) have not been considered in the mask.
As supported by the works of \citet{srm09} and \citet{p12} for linear polarization due to diffusion processes, we adopted 
an equal weight for each atomic line considered in the mask.
This allows very strong detections of linear polarization in each LSD profile. On the other hand, to study the polarization of individual lines required the averaging of  8 $Q$ and 8 $U$ spectra.

In the final eight columns of Table\,\ref{tab1} we give the polarimetric parameters related to two bright spots 
inferred from our measurements (see Section 5.): observed maximum of linear polarization $P_L = \sqrt{Q^2+U^2}$, polarization 
angle $\theta$, position angle $\chi$, and projection angle to disk center $\mu$, as described in Section\,5.

\begin{figure*}
\centering
\includegraphics[width=13cm,angle=270.] {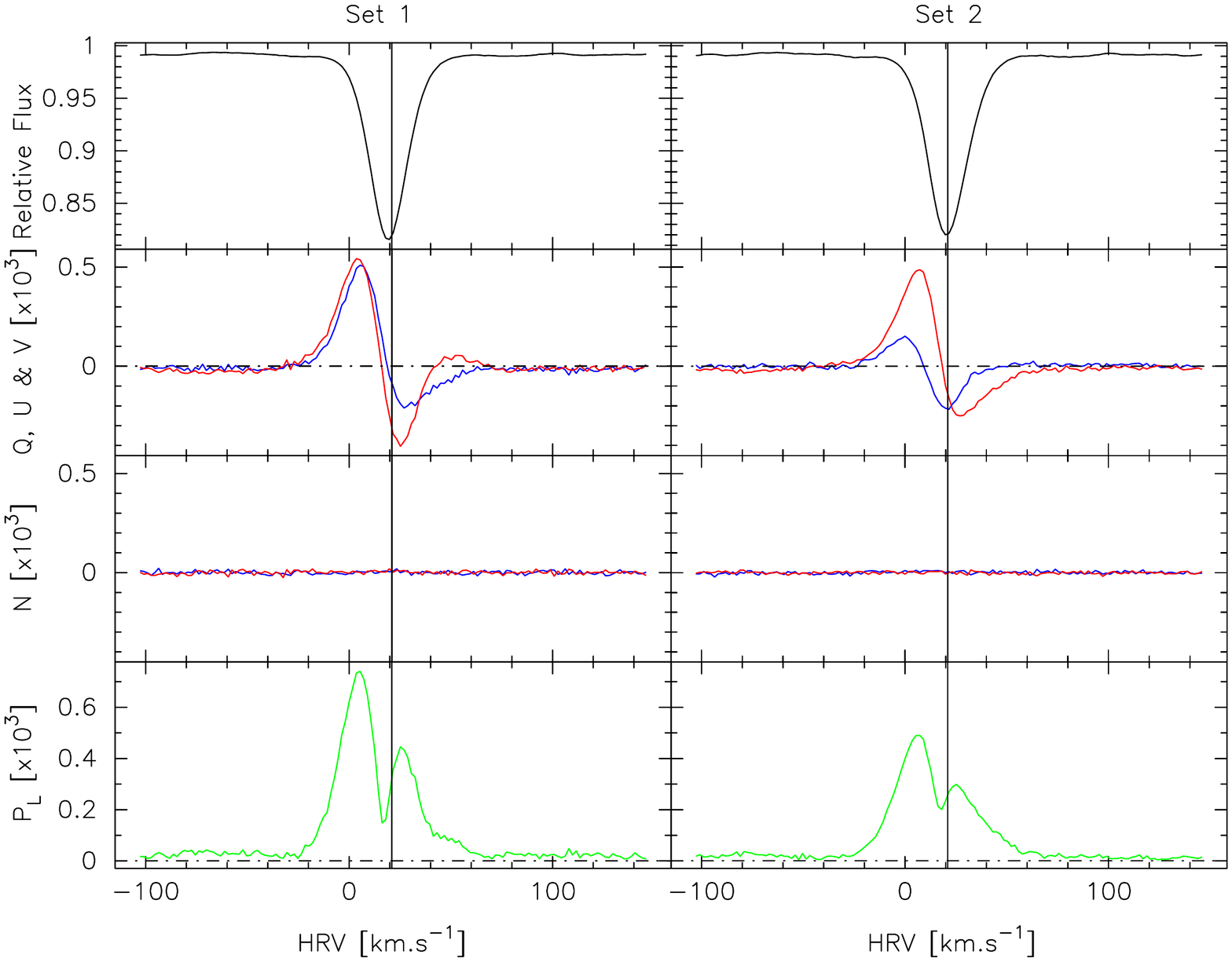}
\caption{Stokes $I,Q,U$ and null polarization $N$ LSD profiles of Betelgeuse for November-December 2013-January 2014 (left, Set\,1) 
and September 2014 (right, Set\,2). 
The upper panel shows $I_{Q,U}$ intensity profiles, the  second panel shows Stokes $QU$ profiles, the third panel shows the null polarization $N_{Q,U}$ profiles and the bottom panel shows the linear polarization $P_L=\sqrt{Q^2+U^2}$ profiles (Blue for Stokes $Q$ and red for Stokes $U$).  
The vertical lines correspond to Betelgeuse's heliocentric radial velocity.}
\label{fig1}
\end{figure*}  

\section{Stellar origin of the linearly polarized spectrum of Betelgeuse}

Figure\,\ref{fig1} shows the averaged LSD profiles obtained in November-December 2013 and January 2014 (hereafter named Set\,1), 
and September 2014 (hereafter named Set\,2). 
Note that both sets consist of the average of eight Stokes $Q$ and $U$ profiles (Table\,\ref{tab1}).
Figure\,\ref{fig1} shows that the Stokes $Q$ and $U$ signals full amplitude are up to the 7-9 $10^{-4}$ level and centered on the mean $I$ profile. 
For clarity, we also show the heliocentric radial velocity of \object{Betelgeuse} (HRV hereafter), about 21\,\kms (Mathias et al., in preparation).
For comparison, the circular polarization Stokes $V$ signal obtained during the same nights 
 is about ten times weaker than the linear polarization \citep[Mathias et al. in preparation, see also][]{adk10}. 
Finally, Fig.\,\ref{fig1} presents, at the same scale, the null polarization profiles $N$: they do not show any signal above 
the noise level of a few  $10^{-5}$. 
Since these profiles (including Stokes $V$) are obtained from different combinations of spectra acquired through the same rhombs but 
with different orientations, it demonstrates that the linear polarization observed in \object{Betelgeuse} 
is not a spurious signal due to Narval. This statement is also supported by the non detection of linear polarization in AGB stars which is described later in this section.

\begin{figure*}
\centering
\includegraphics[width=13cm,angle=270.] {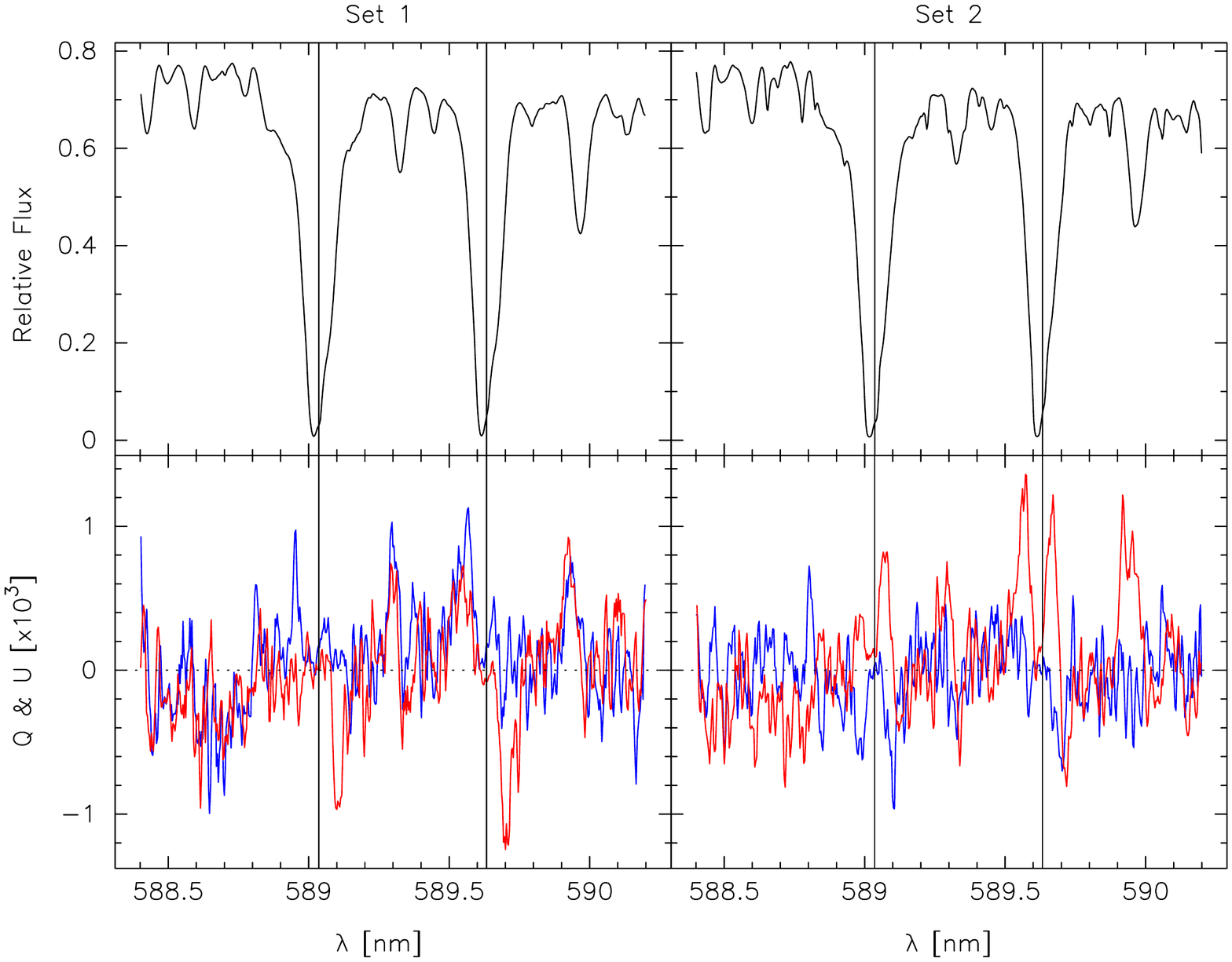}
\caption{Linearly polarized spectrum of Betelgeuse (with the continuum polarization level set to zero) around the \ion{Na}{i} lines D2 (588.9\,nm) and D1 (589.6\,nm) for
Set\,1 (left) and Set\,2 (right). In addition a blend of lines of \ion{Ni}{I} and \ion{Fe}{I} (at 589.3\,nm, between D2 and D1) and a 
line of \ion{Ti}{i} at 589.9\,nm are visible.
The upper spectrum shows Stokes $I$ and the lower spectra show Stokes $Q$ (blue) and Stokes $U$ (red).
The vertical lines correspond to the position of the  \ion{Na}{i}\,D lines in the stellar rest frame.}
\label{fig2}
\end{figure*}

\begin{figure*}
\centering
\includegraphics[width=10cm,angle=270.] {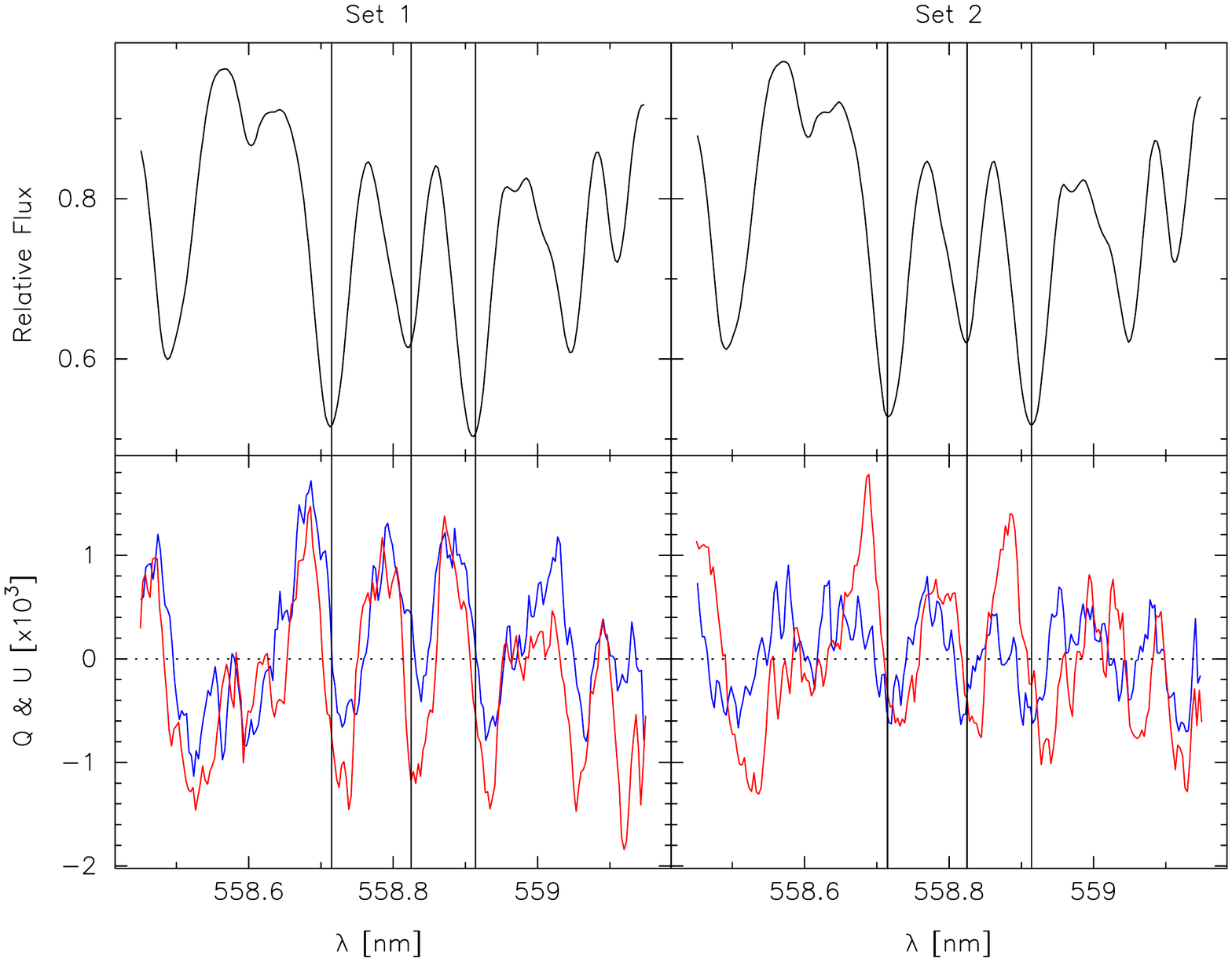}
\caption{Linearly polarized spectrum of Betelgeuse (with the continuum polarization level set to zero) around the \ion{Fe}{i} (558.7\,nm), \ion{Ni}{i} (558.8\,nm) \& 
\ion{Ca}{i} (558.9\,nm) in Set\,1 (left) and Set\,2 (right) .
The upper spectrum shows Stokes $I$ (intensity) and the lower spectra show Stokes $Q$ (blue) and Stokes $U$ (red).
The vertical lines correspond to the positions of the  \ion{Fe}{i}, \ion{Ni}{i} \& 
\ion{Ca}{i}  lines in the stellar rest frame.}
\label{fig8}
\end{figure*}

Figure\,\ref{fig2} illustrates, for the same sets of data shown in Fig.\,\ref{fig1}, the linear polarization spectrum in the region 
of the D1 and D2 \ion{Na}{i} lines (we recall that for all the presented polarized spectra the continuum polarization level is set to zero). 
In addition to the \ion{Na}{i}\,D lines, this figure also shows a blend of lines of \ion{Ni}{I} and \ion{Fe}{I} (at 589.3\,nm, between D2 and D1) and a 
line of \ion{Ti}{i} at 589.9\,nm. 
Other examples of photospheric lines are presented in Fig.\,\ref{fig8}. 
The structure of the individual polarized lines is similar to that of the LSD profiles (the case of the resonance lines of 
\ion{Na}{i} will be studied in Section\,4.2). 
Slow time variations from November 2013 to April 2015 of $QU$ individual and LSD profiles are observed 
(see Section\,6.1 and Fig.\,\ref{fig12}). 
These observations suggest that the polarized signal is related to the lines of \object{Betelgeuse}.

The linear polarization spectrum of cool evolved stars has already been detected using the LSD method in a number of Miras
\citep{laf14,laf15} and RV\,Tauri stars \citep{swl15,laf15}.
In these cases, the polarization seems to be linked to the atmospheric shock wave \citep{flg11} since its shape
changes with pulsation phase, being maximum near maximum brightness, which corresponds to the shock passage through
the atmosphere.
The $Q$ and $U$ profiles change from one Mira star to another and, for a given star, from one cycle to another. 
On the other hand, no linear polarization was detected at the 5 x$10^{-5}$ level on four magnetic, non-pulsating, AGB stars of about the same 
effective temperature (namely \object{$\beta$\,Peg}, \object{15\,Tri}, \object{8\,And} \& \object{$\beta$\,And}) studied with Narval  (L\`ebre et al. in preparation). For these stars, a significant Zeeman Stokes $V$ signal has been observed previously \citep{kac10,kac13}. 

The observations reported above show that the linearly polarized spectrum of \object{Betelgeuse}, as well as that observed in
Miras and RV\,Tauri stars, is not due to a spurious signal from the polarimeters but has a stellar origin.
In addition, all stars for which linear polarization in the photospheric lines has been detected belong to a category of 
cool stars for which linear polarization of the continuum has also been observed.

At this point we can argue that this linear polarization is not of Zeeman origin: it is much stronger than the circular polarization observed in \object{Betelgeuse}, whereas Zeeman linear polarization is typically ten times smaller than the associated circular polarization \citep[e.g.][]{wdl00,rkw13,rkw15}.  We will also show in Section\,4 that the polarization decreases with wavelength, a phenomenon that is also in contradiction with a Zeeman origin.

\section{Interpretation of the linear polarization in the atomic lines of Betelgeuse}

\subsection{Linear polarization of the continuum of Betelgeuse and its depolarization by atomic lines}

\object{Betelgeuse} is known to present linear polarization of its continuum beyond the 0.5\,\% level in the blue 
\citep[e.g.][for the most recent studies]{h84,cs84,nrm94,mn00}.
This is a clear sign of the existence of anisotropy in the radiation field: it is generally believed that the linear polarization 
of the continuum is mainly due to photospheric Rayleigh scattering and Mie scattering by circumstellar dust with an anisotropy induced by one or several 
brightness spots \citep{sc84,cs84,d86,mn00}. 

From our high S/N, high resolution linearly polarized spectra of \object{Betelgeuse}, we observe significant linear polarization signal associated with most of the relatively strong lines.
This behavior is reminiscent of that observed on the Sun and called the `second solar spectrum', except that 
for the latter the anisotropy of the radiation field is due to limb darkening \citep{skg00}. 
The second solar spectrum has two main contributions \citep{sth83}: i) depolarization of the continuum by the lines; 
ii) intrinsic polarization of individual lines. 
While the majority of the lines are expected to depolarize the continuum \citep{fs03}, only a small number of them are 
expected to provide significant intrinsic polarization contributions \citep{sth83,sthb83}. While explaining the origin of the intrinsic polarization of individual lines is the subject of many publications, the case of the depolarization of the continuum is less well documented in the literature. This is because each intrinsic line polarization is unique and has to be explained by atomic physics; furthermore the involved coherent scattering is sensitive to magnetic field through the Hanl\'e effect \citep[e.g.][]{s09} which has been applied to the investigation of weak fields on the Sun. On the other hand the physics involved in the process of line depolarization is the same for all the lines. We propose two levels of interpretation: \citet{sth83} present empirical relationships which show that `the polarization line profiles of depolarizing lines appear to have approximately the same shape as their corresponding intensity profiles'; then \citet{fs03}  present a theoretical analysis of the depolarizing lines; their properties may be also inferred from the work of \citet{ll04} from which it can be derived that the depolarization is mainly due to absorption and spontaneous re-emission of any photon by the atom at the origin of the spectral line: this is an incoherent process which emits an unpolarized photon, whatever the incident photon is polarized or unpolarized. In the present work the process that we call `depolarization of the continuum' of \object{Betelgeuse} is the same as that described above for the Sun.

In all our observations of \object{Betelgeuse} the \ion{Na}{i}\,D lines appear strongly polarized and D1 is as polarized,
or even more polarized, than D2 (see Set\,2 of Fig.\,\ref{fig2}).
However, the intrinsic polarizability of the D1 line is much smaller than that of D2 \citep{ll04}.
The strong polarization of D1 in the spectrum of \object{Betelgeuse}, comparable to that of D2, cannot be intrinsic. 
We propose that it originates from the depolarization of the continuum. 
Since in our Narval data we set the (unmeasured) continuum polarization level to zero, any depolarization appears 
as a net signal, as observed and will be illustrated later in Section\,5.3 . 
Next to D1, D2 also depolarizes the continuum and, in addition to that signal, the intrinsic polarization of this line is contributed. 
This supports the notion that the signals in the two lines are not identical. 
Despite these differences, the similar signal amplitudes in D1 and D2 points to the fact that depolarization dominates the 
signal in D2. 
From this observation, we can generalize and assume that the linear polarization signal present in the photospheric lines 
of the spectrum of \object{Betelgeuse} is dominated by the depolarization of the continuum.

Figure\,\ref{fig2} shows that metallic lines in the region of D1 and D2  are polarized (see Section 3), having the same shape than the LSD profiles presented in Fig\,\ref{fig1}. 
Figure\,\ref{fig8} shows for the same two dates an extraction of the linearly polarized spectra around three lines 
(\ion{Fe}{i} 558.7\,nm, \ion{Ni}{i} 558.8\,nm, \ion{Ca}{i} 558.9\,nm) which are also known to depolarize the continuum of the solar 
spectrum \citep{g00}. 
We see that all the prominent lines visible in this region are polarized and present shapes similar to that of the LSD 
profiles in Fig.\,\ref{fig1}.
\citet{sth83} show that in the Sun "most spectral lines have a very small intrinsic 
polarization, their main effect being to depolarize the continuum polarization".
For the depolarizing lines, the shape of the polarized profiles appears to be statistically identical for all
absorption lines \citep[relationship 3.11 of][]{sth83}. 
Then this common shape would lead to an averaged profile in the LSD procedure of the $Q$ and $U$ profiles of \object{Betelgeuse}, and would explain why we obtain a very significant signal. 
However in the case of \object{Betelgeuse}, the linearly polarized profiles of the lines are much more complex than in the case of the Sun and in general they do not mimic the intensity profiles. Actually, they appear to present shapes varying with time, and this complexity is due to the formation process  of the polarization of the continuum (see Section 5.).
Hence, depolarization of the continuum seems to be an important contribution to the linearly polarized spectrum of 
\object{Betelgeuse}.

\subsection{Properties of the linearly polarized spectrum}

In order to investigate the hypothesis of depolarization of the continuum as the origin of the linearly polarized spectrum of Betelgeuse, we have studied some properties of its LSD averaged profiles (as introduced in Fig.\,\ref{fig1}), making sub-masks selected for their mean depth or their mean wavelength, from a mask similar to the one described in Section\,2 but including many more atomic lines (all lines with a central depth greater than 0.01 of the continuum). 
Depending on the required conditions, the number of lines in a given sub-mask is between 2\,000 and 15\,000.
Changing the number of lines showed that this parameter does not significantly affect the $QU$ profiles obtained using the sub-masks.
We have checked the variations of the strength of the Stokes $Q$ and $U$ signals and of the linear polarization 
$P_L = \sqrt{Q^2+U^2}$ with respect to depth and wavelength of the lines and found strong correlations.
In the following, we concentrate only on data from Set\,1, but fully compatible results have also been
obtained for Set\,2.

\begin{figure}
\centering
\includegraphics[width=6.5cm,angle=-90.] {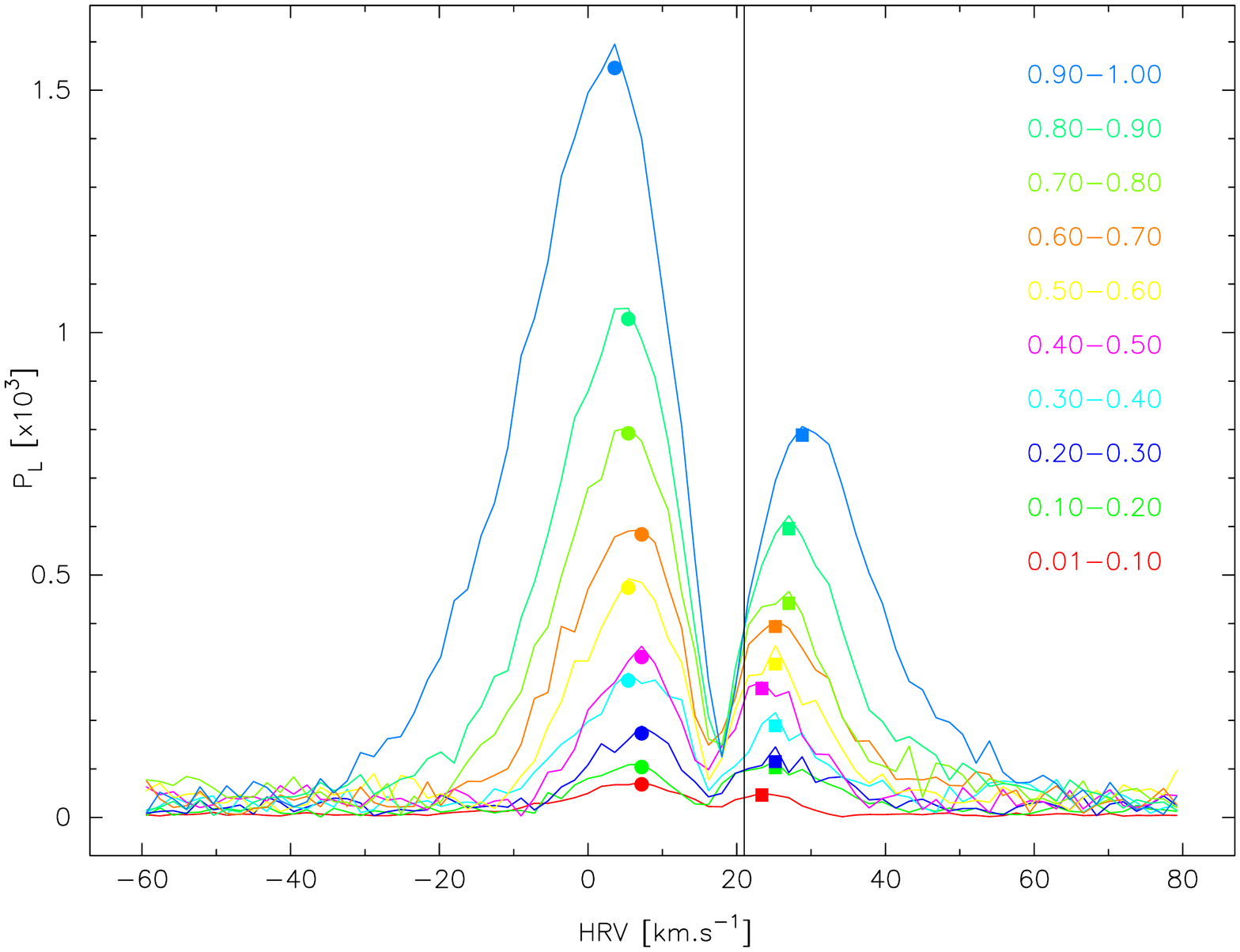}
\caption{For Set\,1, variations of the linear polarization $P_L$ of Betelgeuse from the LSD profiles with the depth of the lines. 
Ten sub-masks are used with line depth domains as indicated on the graph.
The vertical line corresponds to Betelgeuse's heliocentric radial velocity.}
\label{fig3}
\end{figure}

In order to investigate the dependence of the linear polarization $P_L$ with the depth of the lines, we have made ten 
sub-masks, restricting the 
individual spectral intensity line depths from 0.01-0.1 of the continuum (the weakest lines, i.e., those mainly formed in the lower part of the atmosphere) to 0.9-1.0 of the continuum (the strongest lines, formed higher in the atmosphere). 
Figure\,\ref{fig3} shows the mean polarization profiles and  
the maxima of $P_L$  (i.e., main peak maxima, represented by dots for the blue lobe and squares for the red lobe) for the 10 mean depths: the polarization strongly increases with line depth, that is, with altitude. We checked the mean wavelength of the ten sub-masks and found that they were included in a 100 nm range, meaning that they did not differ significantly. Also, in Fig.\,\ref{fig3} the velocities corresponding to the maxima of $P_L$ deviate more and more from the photospheric velocity (quoted in the Figures as Betelgeuse's heliocentric radial velocity) with increasing line depth (i.e. height in the atmosphere). We interpret this trend as due to the expansion of the atmosphere which will be discussed again  in Section 5. We obtain similar results for both lobes.

\begin{figure}
\centering
\includegraphics[width=6.5cm,angle=-90.] {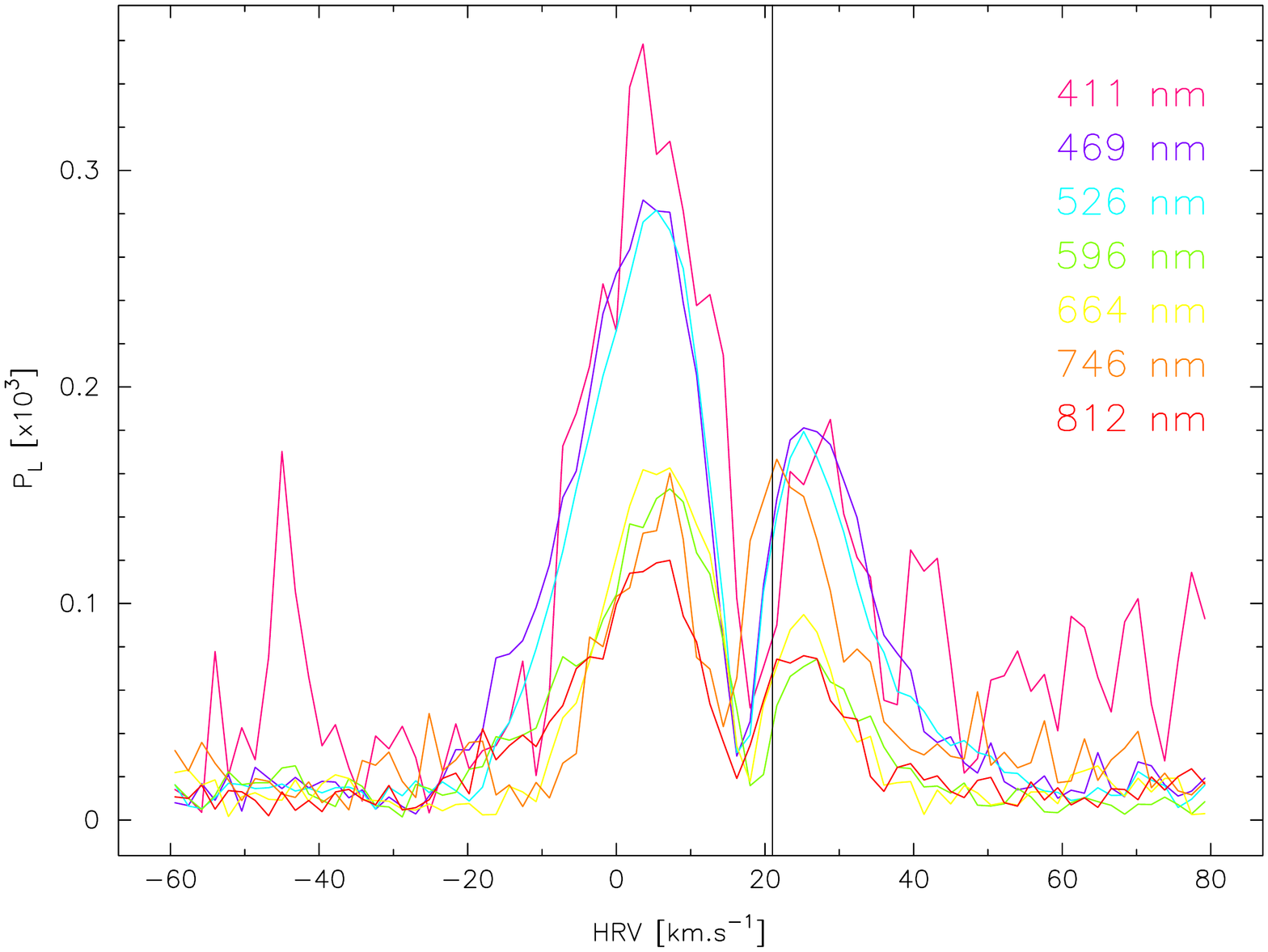}
\caption{For Set\,1, variations of the linear polarization of Betelgeuse $P_L$ along the LSD profiles with wavelength: 
different colors illustrate polarization decreasing when measured with sub-masks corresponding to the mean 
wavelength written on the graph (in units of nm).
The vertical line corresponds to Betelgeuse's heliocentric radial velocity.}
\label{fig5}
\end{figure}

In addition to this variation of the linear polarization with line depth, we also searched for a wavelength effect.
Indeed, the linear polarization of the continuum in \object{Betelgeuse} is known to increase steeply towards 
the blue \citep[e.g.][]{cs84}, almost as steeply as the $\lambda$$^{-4}$ law. 
Figure\,\ref{fig5} illustrates the variations of $P_L$ corresponding to seven sub-masks with wavelength-bandpasses of
350-420\,nm, 420-490\,nm, 490-560\,nm, 560-630\,nm, 630-700\,nm, 700-770\,nm, and 770-1050\,nm. 
The corresponding average wavelengths are given in the figure. The $P_L$ measurements were normalized to a common mean depth of the Stokes I profiles (that corresponding to the reddest sub-mask).

\begin{figure*}
\centering
\includegraphics[width=19cm,angle=0.]{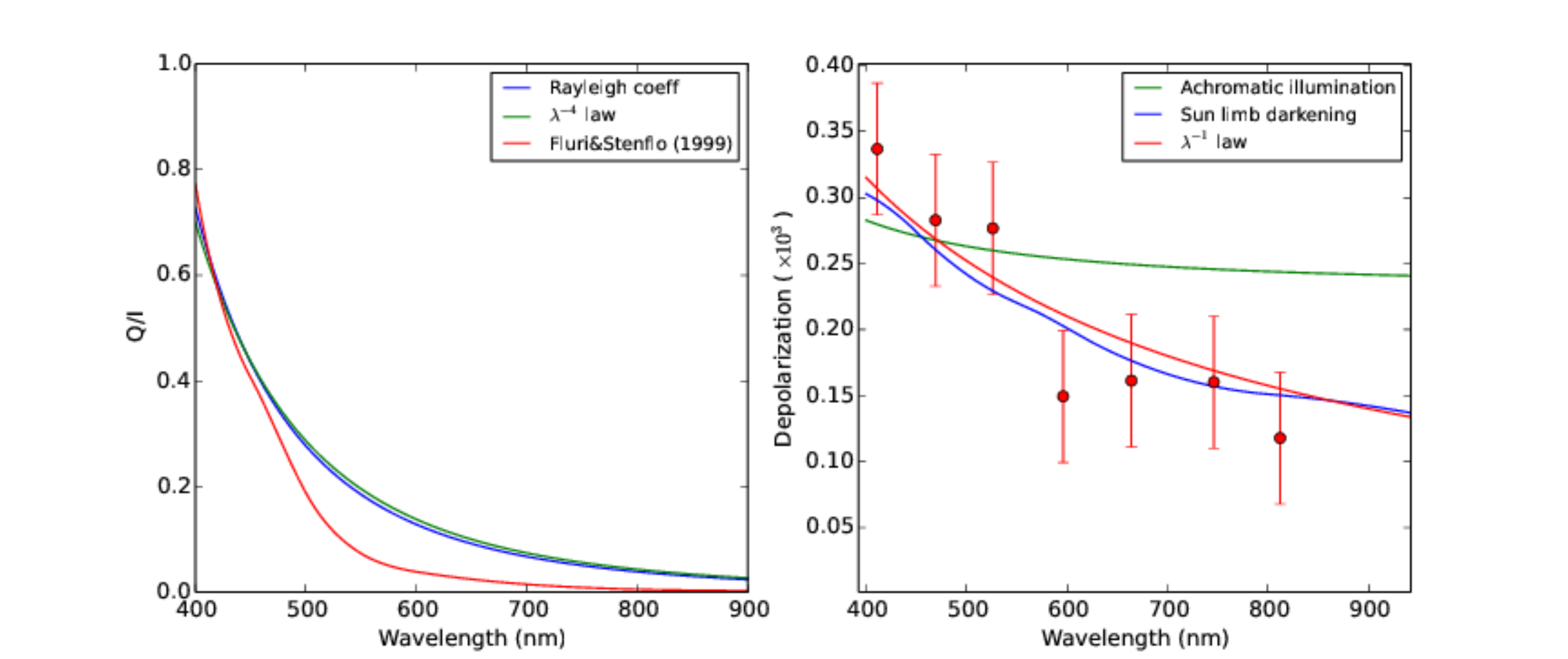} 
\caption{Wavelength dependence of the polarization of the continuum (left) and depolarization in the spectral lines (right) under different 
approximations. On the left are shown the polarization of simple Rayleigh scattering by hydrogen atoms (blue), the approximation $\lambda^{-4}$ (green), 
and the empirical law from \citet{fs99} for the continuum of the Sun taking into account the chromatic dependence of limb darkening (red).
On the right is shown depolarization in the lines,  without taking into account the chromatic variation of limb darkening (achromatic illumination, green curve), 
or including this variation as in the solar case (blue). 
Points are the maxima of the main polarization peak represented in Fig.\,\ref{fig5} with their error bars, together with the power law $\lambda^{-1}$ (red).
}
\label{fig9}
\end{figure*}  

Figure\,\ref{fig5} presents the variation of the linear polarization  with wavelength accross LSD profiles, and the decrease from blue to red is confirmed. However we find that it is consistent with a $\lambda$$^{-1}$ law. In order to interpret this phenomenon, we explored both the diffusion and limb darkening mechanisms.

Figure\,\ref{fig9} (left panel) presents the computation of the expected wavelength dependence of Rayleigh scattering by hydrogen atoms 
as $\lambda_0^4/(\lambda_0^2-\lambda^2)^2$ \citep{k24}, where $\lambda_0$ is the Lyman\,$\alpha$ wavelength 
($\lambda_0=121.6$\,nm), and its usual $\lambda^{-4}$ approximation which closely matches it (blue and green curves, respectively). 
However, this basic wavelength dependence of Rayleigh scattering is modified in stellar atmospheres by the chromatic variation
of limb darkening. 
The red curve in the left plot shows the case of the Sun as studied and described by \citet{fs99}. 
Therefore, due to limb darkening, the polarization of the continuum in the solar case can fall faster than the 4$^{\rm th}$ power of wavelength.

In the right plot of Fig.\,\ref{fig9}, we followed \citet{ll04} to simulate the wavelength 
dependence of this depolarization signal in the lines. 
It is proportional to the Rayleigh coefficient $\lambda_0^4/(\lambda_0^2-\lambda^2)^2$, as is the continuum, but inversely 
proportional to the line opacity and
 the emission coefficients of the lines. 
These three combined contributions result in the green curve labelled achromatic illumination
since we have not considered any wavelength dependence in the anisotropy of the light due to limb darkening. 
The blue curve of this right plot (Fig.\,\ref{fig9}) introduces limb darkening, again for the solar case which is used here as an example and a first approximation. 
We note that in both approximations, with and without chromatic limb darkening, the decrease of the depolarization
amplitude in the lines is moderate. 
For each wavelength we have also represented on the same plot the values of the different maxima of the main peaks shown in
Fig.\,\ref{fig5} and an estimate of their errors.
Superposed to these points we plot as a reference the $\lambda^{-1}$ law (red curve), that
qualitatively reproduces the wavelength dependence of our measurements of \object{Betelgeuse}, hence the depolarization signal. 
Given that limb darkening in \object{Betelgeuse} and in the Sun are expected to be similar since the main 
opacity source is the same ($H^-$), one thus expects a dependence of the depolarization with wavelength 
of the order of $\lambda^{-1}$. 
A more detailed analysis of the line formation conditions in the atmosphere of \object{Betelgeuse} would likely give a better fit
to this observation, but it is beyond the scope of this paper.
A very interesting consequence is that such observations would provide new constraints on the limb darkening 
of \object{Betelgeuse}. Our present investigation shows that the  difference of the wavelength dependence observed for the continuum \citep[about $\lambda$$^{-4}$ law, e.g.,][]{cs84} and the lines (about $\lambda^{-1}$ law) can be explained if we deal with depolarization of the continuum in the lines, taking into account the diffusion and limb darkening mechanisms.

In summary, the linear polarization observed in the atomic lines of Betelgeuse appears to be due to the depolarization 
of the continuum. 
\object{Betelgeuse} is the first star in which the depolarization of the continuum is observed in the \ion{Na}{i} D1 line, 
since it is neither observed in the Sun \citep{sgk00} nor in the Mira star \object{$\chi$\,Cyg}, nor in the 
RV\,Tauri star \object{R\,Sct} \citep{laf15}. 
Also, no prominent intrinsic polarization is observed for the \ion{Na}{i}D lines. However the polarized profiles are different for D1 and for D2 both in Stokes $Q$ and $U$. As stated above, it is believed that strong polarization in D1 can only originate from depolarization of the continuum, so we can infer that in D2 we also deal with intrinsic polarization which cancels, in part, the effect of depolarization of the continuum. The polarization difference can reach about  10$^{-3}$ whereas the intrinsic polarization of D2 observed in the polarized solar spectrum is 3\,x10$^{-3}$ \citep{bl09}. 

\section{An analytic model for mapping bright spots at the surface of Betelgeuse using its linearly polarized spectrum}

The analysis of our spectra of \object{Betelgeuse} described in Section\,4 shows that we are mainly 
observing the depolarization of the continuum. In this section we use the averaged $Q$ and $U$ LSD profiles.  Although they do not provide the local polarization of the continuum, the information contained in these spectra allows the inference of the surface location and relative brightness of the bright spots inducing anisotropies in the radiation field. 

\subsection{Polarization of the continuum of Betelgeuse}
The observed polarization of the continuum may be explained as a combination of Rayleigh and Mie scattering \citep[e.g.][]{d86}. 
However, in order to depolarize the continuum, the observed photospheric lines must be formed at or above the height at which the 
continuum itself becomes polarized. 
Mie polarization by molecules or dust above the photosphere is thus formed too high and can be excluded from our picture.  
This does not mean that there is no Mie polarization in the observed continuum, 
but just that Mie scattering takes place farther from the star,  above where the photospheric lines have formed and have 
depolarized the continuum, which at this point can only carry polarization due to Rayleigh scattering. 
Since the average line depolarization is of a few times $10^{-4}$ of the continuum flux, this is the minimum contribution of Rayleigh scattering
to the total polarization of the continuum.

Scattering polarization is linear, and orthogonal to the scattering plane. 
At any position angle of the disk of Betelgeuse, the polarization due to scattering will be at a tangent to the local limb, 
orthogonally to photons coming from the deep photosphere and scattered towards us. 
As we change azimuth, the plane of polarization will rotate. 
If the star at the height of its photospheric layers were a perfect sphere and if the amount of polarization were azimuthally symmetric,
the integration over the stellar disk would amount to zero and no depolarization signal would be detected.
As a consequence, two possibilities appear:  \object{Betelgeuse} is not a sphere, and/or its brightness is not azimuthally symmetric. 
Most observations and analysis conclude that \object{Betelgeuse}  
conserves a spherical symmetry at the height associated with our observations \citep[e.g., observations by SPHERE, ][]{ker15}.
We are left with the second possibility: the locally-induced polarization is not azimuthally symmetric. 
There are several manners in which the azimuthal symmetry may be broken, among them are:
clouds over the star that only let us see some regions of the lower atmosphere; plumes 
of material rising high above the photospheric level with higher radiation anisotropy and higher polarization level; dark regions of the photosphere emitting less light and hence inducing less polarization; brighter, hotter regions of the photosphere emitting more light and  more polarization.
In the following section, we adopt the latter explanation for its simplicity and since bright spots have been observed by interferometry on \object{Betelgeuse} \cite[e.g.][]{hpl09} and on \object{Antares}  \citep{thb97}. In addition, numerical simulations \citep{chi09,chi10,chi11} also indicate that bright convective spots are expected in the atmospheres of red supergiants.

\subsection{A two spots hypothesis for interpreting the polarized spectrum}

We assume the presence of bright spots in the deep photosphere of Betelgeuse, located at particular positions over the disk.  
We suppose in the following that the (Rayleigh) scattering matter is located vertically above these spots.
The observed LSD profiles shown in Fig.\,\ref{fig1} are composed of two lobes, a blue one (of positive polarity in both
Stokes $Q$ and $U$ in both Sets), and a red one (of negative polarity in both Stokes $Q$ and $U$, also for both Sets),
with the blue lobe being in general stronger than the red one.
With respect to the center of the intensity mean profile, 
the two lobes are located respectively at about +5 and +25\,\kms\  $HRV$ for both sets.
To explain this behavior we consider the simple scenario of two photospheric bright spots referred to above and
in Section\,4: one spot would be related to the structures that move toward the observer (blue features, spot1 hereafter),
the other would be linked to structures moving away (red features, spot2 hereafter).
Actually, the observed radial velocities are the combinations of those of the layer where the lines are formed in the
deep photosphere and that of the layer higher in the atmosphere where the light is scattered (because of a very long rotational period \citep[e.g.,][]{u98}, contributions from the rotational velocity of the star are negligible).
Both layers are able to contribute to velocities that can be blueshifted or redshifted with respect to the central part of the line:
ascending versus sinking elements exist in the photosphere, and we assume the outer layers of Betelgeuse are in expansion. In front
of the disk the expanding shells are blueshifted, whereas beyond the limb the expanding matter is redshifted.
The observed intensity profiles are consistent with a model in which radiation from the stellar photosphere is scattered and reemitted by an expanding envelope(s) of radius larger than the photosphere radius \citep{grt75}. Therefore, the blue component of the LSD polarimetric profile could be the consequence of scattering by the approaching half of the expanding envelope, while the corresponding red component could be due to the scattering of the same expanding envelope, radiated from its receding half. The expansion of the atmosphere is illustrated in Fig.\,\ref{fig3} where the strongest lines, formed higher in the atmosphere, present larger expansion velocities.

\subsection{An analytic model for mapping bright spots at the surface of Betelgeuse}
\begin{figure*}
\centering
\includegraphics[width=15.cm,angle=0.]{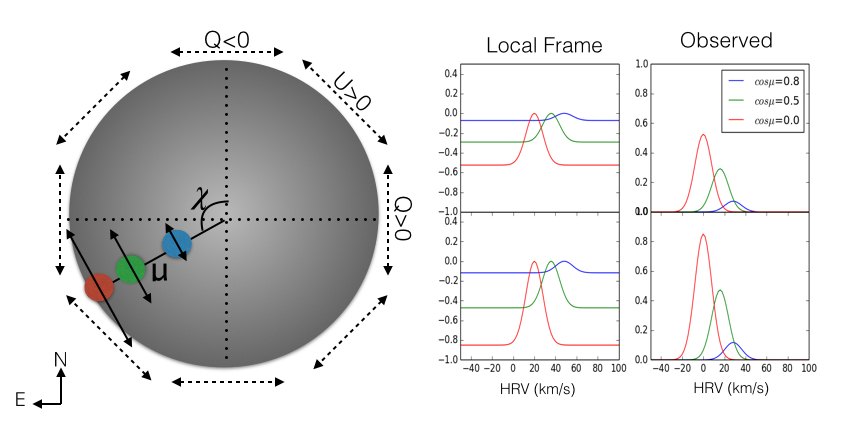} 
\caption{Coordinates on the disk of Betelgeuse and polarized spectra in the case of one bright spot. In the left of the figure a cartoon of the stellar disk shows the polar coordinates for three bright spots at the same $\chi$ but at different $\mu$. Accross each spot the arrow shows both the orientation and relative polarization expected. Around the disk the dashed arrows show the definition of the Stokes parameters $Q, U$. In the right-hand part of the figure, the leftmost two plots show Stokes $Q$ (top) and $U$ (bottom) for the three spots as they would be seen in a resolved stellar disk. The two rightmost plots show what would be observed with Narval and the radial velocity scale is  shifted to the $HRV$ of \object{Betelgeuse}. See Section 5.3 for a complete description.}
\label{fig15}
\end{figure*}

Following this scenario, we can propose an analytical model that links the observed polarizations to the positions and relative brightness
of the bright photospheric spots on the visible stellar disk. The two polar coordinates that we derive from the polarimetric data are the position angle $\chi$ and the projection angle $\mu$. Figure\,\ref{fig15} presents what is observed in the case of one bright spot. In the left of the figure, a cartoon of the stellar disk shows the polar coordinates for one bright spot at three locations, with the same position angle $\chi$ but at different projected distances $\mu$ to the disk center. On top of each spot the arrow shows both the orientation and the relative polarization expected from the spot alone. Around the disk the dashed arrows show the definition of the Stokes parameters $Q, U$ as measured in the spectral lines and described in Section 5.3.1. In the right-hand part of the figure, the leftmost two plots show Stokes Q (top) and U (bottom) for the three spots as they would be seen on a resolved stellar disk. The continuum is polarized and the lines depolarize at the wavelength corresponding to the projected radial velocity. The two rightmost plots show what would be observed with Narval: the continuum polarization is set to zero and what was a depolarization signal becomes a net signal over zero. The radial velocity scale is also shifted to the $HRV$ of \object{Betelgeuse}.

\subsubsection{Getting the position  angle $\chi$ from our Narval data}
First, each spot will lead to an excess of polarization perpendicular to the radius joining the center of the disk of 
\object{Betelgeuse} with the spot, this radius forming an angle $\chi$ with the North-South meridian of the disk, 
its position angle. 
With respect to this North-South meridian we expect an excess of polarization with orientation $\theta$, the polarization angle.  
Conversely, observing a polarization at $\theta$ position, we can infer $\chi$ as $\theta \pm$ 90$^\circ$ that is, an ambiguity 
of $180^{\circ}$. 
Using the classical linear polarization relations, as described by \citet{bll09}, and taking into account that with Narval 
(as for ESPaDOnS), positive $Q$ is defined in the North-South direction and positive $U$ is defined at 45$^\circ$ (counterclockwise) relative to positive $Q$, the 
polarization angle is given by
$$\theta = 0.5\arctan({U/Q}),$$
(if $Q > 0$ and $U \geq 0$ ;+180$^\circ$ if $Q > 0$ and $U < 0$; +90$^\circ$ if $Q < 0$).

The position angle is therefore
$\chi = \theta \pm 90^{\circ}$. In the next sections we choose the option  $\chi = \theta + 90^{\circ}$. This choice will be discussed in  Section 6.2. 
Figure\,\ref{fig12} shows the LSD profiles of Stokes $Q$ and $U$ and $P_L$ for our 12 observations. 
From the relationships above we can determine the maxima of linear polarization $P_L$ and its associated $\theta$, from which we can infer $\chi$. 

In order to map the spots at the surface of \object{Betelgeuse} we now compute another coordinate in addition to $\chi$, the  projection angle $\mu$ from the disk center, and brightness ratios among the different spots.

\subsubsection{Projection angle $\mu$ and brightness ratios}
 To infer the  projection angle $\mu$ we make the approximation that all photospheric motions and the expansion velocity of the 
scattering layer are radial and combine at the polarization level to an identical velocity $V_0$. We  represent the velocities as in the case of an expanding atmosphere (as referred to in Section 5.2). Then the blueshifted motions are related to the visible (approaching) hemisphere of the star and redshifted signals are related to the opposite (receding) hemisphere of the star.
As we explain below we assign an ad hoc value $V_0 = 50$\,\kms. 
The polarization of a spot at a projection angle $\mu$ from the disk center will acquire a line-of-sight velocity of
$V = V_{0}\,\cos\mu$. 

The sign of $V$ is negative or positive respectively for blueshifted or redshifted motions with respect to the
$HRV$ of \object{Betelgeuse} (about 21\,\kms). 

By fixing $V_0$ we can determine $\mu$ on the visible/opposite hemisphere from the $HRV$ position of each observed 
polarization lobe. 
Having thus determined $\chi$ and $\mu$, we have the coordinates of the spots on the disk. 

The total polarization can be approximated to be due exclusively to the intrinsic brightness $B$ of the
photospheric spot, corrected by the angular dependence of Rayleigh scattering. 
The exact dependence of the depolarization signal on the brightness is complex. To simplify the model we assume that
all other contributing factors apart from brightness and position will be identical from one spot to
the other. 
Hence we can deduce brightness ratios among the different spots by assigning the brightest one an ad hoc brightness $B=1$.

Hence $P_L=\sqrt{Q^2+U^2} \propto B \sin^2\mu$

\subsection{Mapping the bright spots}
The above computations can be generalized at every single 1.8\,\kms \,velocity bin along the Stokes $Q$ and $U$ profiles, enabling us to derive a map of the involved 
bright spots on \object{Betelgeuse}.

\begin{figure*}
\centering
\includegraphics[width=13.cm,angle=270.]{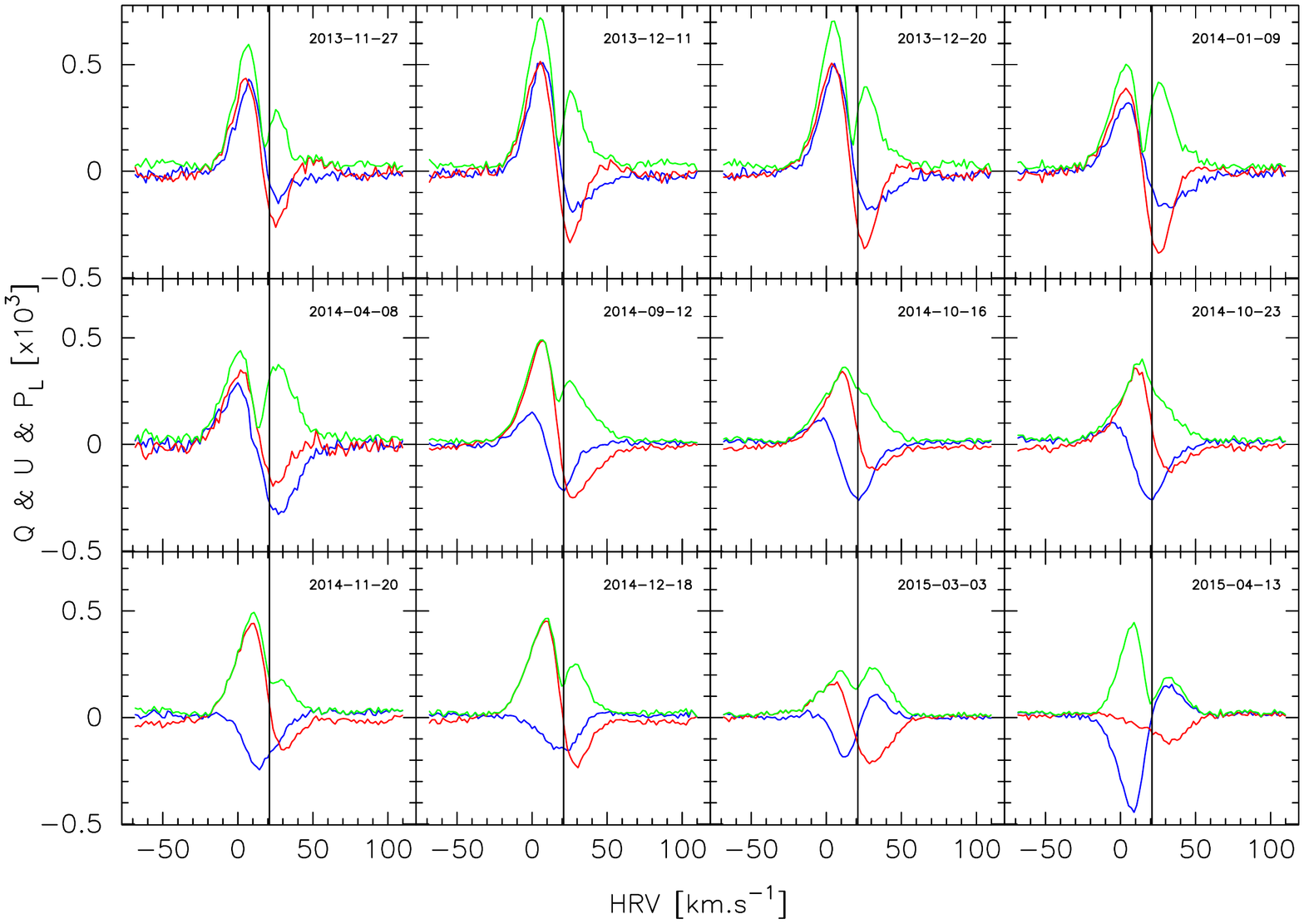}
\caption{Stokes $Q$ (blue) Stokes $U$ (red) and $P_l$ (green) LSD profiles of photospheric lines of \object{Betelgeuse} from November 2013 to April 2015 
for the 12 dates given in Table\,\ref{tab1}.
The vertical lines correspond to Betelgeuse's heliocentric radial velocity.}
\label{fig12}
\end{figure*}

In the case when the peaks in $Q$ and $U$ are at the same velocity, as during 2013 and in January 2014, the peak 
of $P_L$ coincides with them (e.g., as shown in Fig.\,\ref{fig12}).
However in general the peaks are not aligned. The model will interpret this non-coincidence as an extended spot spanning from the position of the peak in $Q$ to the 
position of the peak in $U$. 
This is due to a combination of stellar phenomena broadening spectral lines, along with the intrinsic spectral resolution 
of our instrument. 
In order to take into account the wings of these broadened signals with our model, we must interpret not only 
the peaks but also the wings of these broadened profiles, implying that we must use a value of $V_0$ which 
encompasses both the physical velocities in the photosphere and the broadening of the profiles.
To determine the latter, we fit a Gaussian to each of the observed lobes in $Q$ and $U$. 
The standard deviation of these Gaussians was found to be between 5 and 10\,\kms\ in roughly 75\,\% of the cases. 
To take into account this width we have to  set a value of $V_0$ = 50\,\kms.
This is not a physical value for the velocity of the  photosphere, but rather only a numerical value that allows us to 
implement our model including the broadening of the signal by our instrument and by stellar
phenomena neglected in our model. 
We conducted several trials and saw that the result of the mapping does not depend much on the value of $V_0$.
We can apply the model to all our data and map the spots at the surface of \object{Betelgeuse}. 
This can be seen in Fig.\,\ref{fig13} where one map per observation date is shown. For each observation date, Fig.\,\ref{fig13} shows two planispheres (with the same orientation on the sky): the left one shows spot1 (corresponding to blueshifted motions), the right one spot2 (corresponding to the redshifted motions). In our present interpretation, the redshifted signals corresponding to spot2 arise from the invisible hemisphere of \object{Betelgeuse}. In the case of an interpretation with the redshifted signals being of photospheric origin, the spot would have the same $\chi$ and $\mu$ coordinates but would be on the visible hemisphere.
The model automatically maps the two spots inferred from the profiles and for which some derived characteristic 
numbers are listed at the end of Section\,2 and presented in Table\,\ref{tab1}. In general two $P_L$ maxima are determined (as seen in Fig.\,\ref{fig12} and Fig.\,\ref{fig13}). In this case the numbers in Table\,\ref{tab1} are the outputs of the model. In the case of 16 and 23 October 2014, when the two $P_L$ maxima appeared to merge , the numbers were calculated `by hand'.

\section{Variations of the linear polarization with time and evolution of the spots at the surface of Betelgeuse}

\subsection{Variations of the linear polarization and of the modeled spots}

\begin{figure*}
\centering
\includegraphics[width=\textwidth,angle=0.]{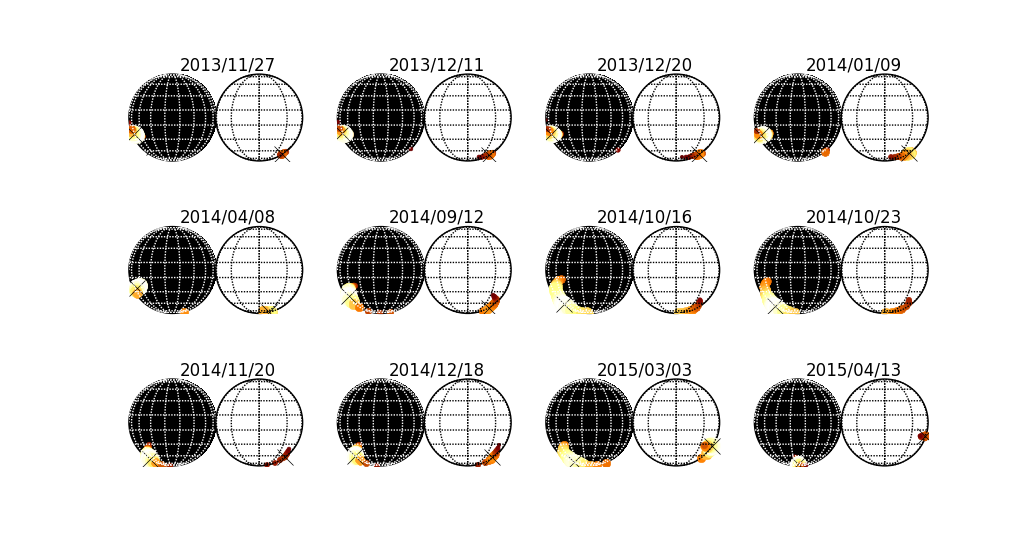} 
\caption{Sequence of images of Betelgeuse from linear polarization for the 12 dates given in Table\,\ref{tab1}. For each date the image (relative intensities) on the left represents the blueshifted signals (spot1) which are located on the visible hemisphere; the image on the right represents the redshifted signals (spot2) which are hypothesized to correspond to the opposite hemisphere. The crosses show the positions corresponding to the maxima of $P_L$. North is up and east is left for all images. The model used is described in Section\,5.}
\label{fig13}
\end{figure*}

As described in Section\,2 and Table\,\ref{tab1}, we observed \object{Betelgeuse} on 12 nights between November 
2013 and April 2015. 
Table\,\ref{tab1} gives the variations of the parameters defined above.  
This shows that the blue-shifted lobe (subscript 1 in Table\,\ref{tab1}) is more polarized than the red-shifted lobe
(subscript 2 in Table\,\ref{tab1}). 
For both lobes the polarization is relatively stable during the time-span, but with some decrease. 
Between November 2013 and January 2014 the polarization angles $\theta_{1,2}$ of the lobes were stable, which we interpret as a stable position 
angle $\chi$ of the spots. Then $\theta_1$ and $\chi_1$ increased smoothly, corresponding to a 
motion of spot1 towards the South. After 2015, the motion of spot1 accelerates and spot1 reaches 
the south pole in April. During this same period, spot2 moves the same angle towards the north. These motions of the spots are illustrated in Fig.\,\ref{fig13}.
  
Changing shapes of the spots are seen on the maps. 
The two main sources of errors that we consider as intrinsic to the model are the already-mentioned PSF broadening 
of the profiles (with standard deviation of 5-10\,\kms) and the photon noise over the profiles (measured to be 
of the order of $\sigma_{ph} = 2 \times 10^{-5}$ of the total intensity over the LSD profiles). It appears, in the end, that our spatial resolution at the surface of the star is roughly  $10\times 10$ square degrees. 
In Fig.\,\ref{fig13} we have made the size of our markers equal to this error.

With respect to linear polarization measurements in the continuum, our geometric model allows us to disentangle several 
spots with different $RV$ and to infer their positions on the disk of \object{Betelgeuse}. However, it does not provide the 
local polarization of the continuum  since we do not know the depolarization coefficient of the lines. 
 An order of magnitude can be derived for averaged lines. Published polarization measurements of the continuum of \object{Betelgeuse} are dominated by Mie polarization \citep[e.g.][]{h84,cs84,ker15}. Predictions for the Rayleigh contribution \citep{d86,jla15} give about $0.3-1$\,\% at maximum in the blue.  
The order of magnitude of the depolarization factor by the lines is therefore about 0.1 in the blue.
We show in Section\,3.2 that the depolarization varies with wavelength.

\subsection{Comparison with quasi-simultaneous observations of VLTI/PIONIER and with older polarimetric observations}

\begin{figure*}
\centering
\includegraphics[width=12.cm,angle=0.]{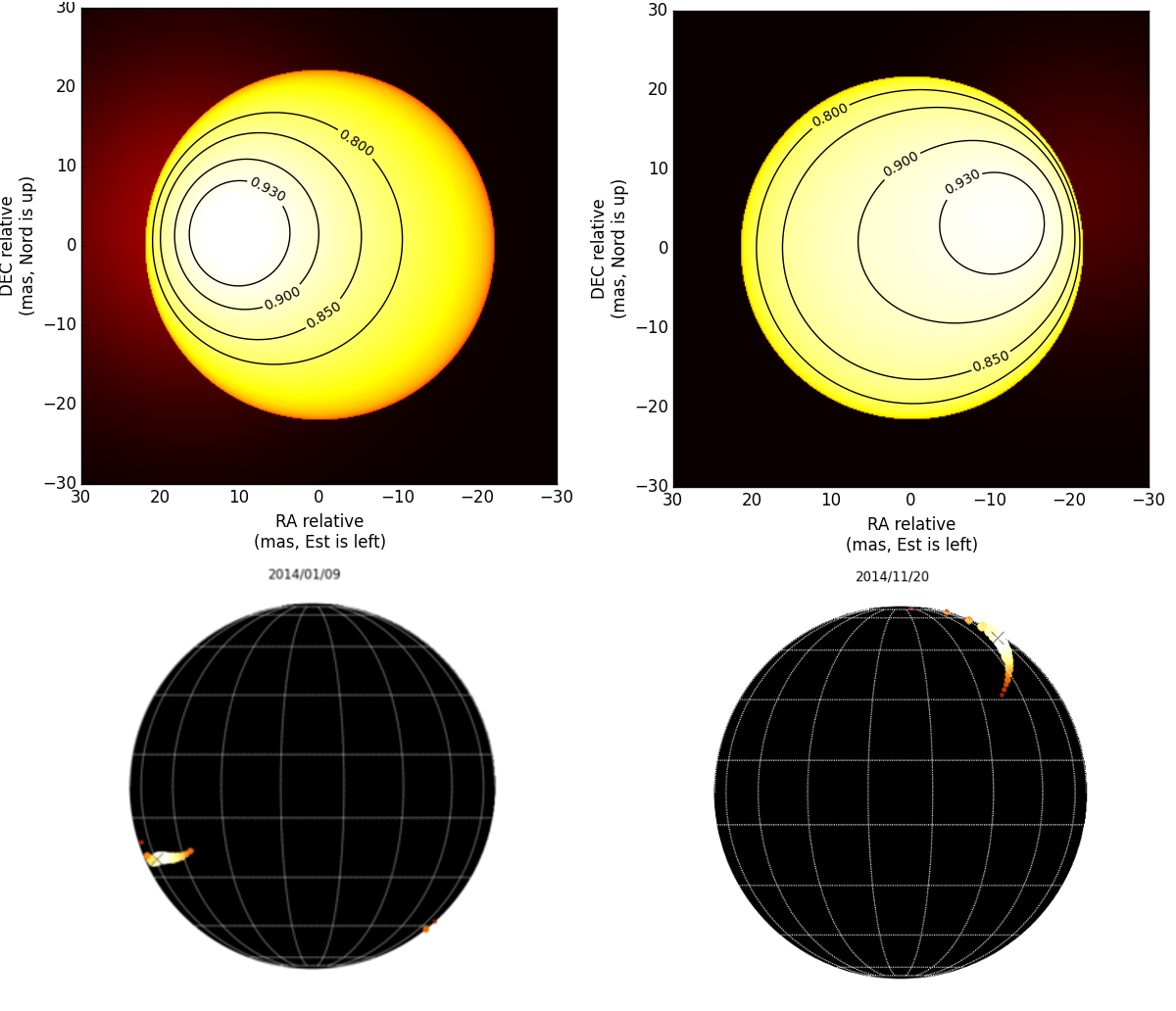} 
\caption{Comparison of intensity images of Betelgeuse obtained with VLTI/PIONIER \citep[upper images, from][]{mkp15} and with TBL/Narval (lower images) during the quasi-simultaneous observations of 2014: January (left) and November (right). The images from polarimetry correspond to spot1 (from Fig.\,\ref{fig13} in January, with $\chi$ + 180$^\circ$ in November). North is up and east is left for all images.}
\label{fig14}
\end{figure*}

\object{Betelgeuse} has been surveyed once per year with VLTI/PIONIER since 2012 \citep{m14,mkp15} and a large hot spot has been resolved near the limb. This hot spot appears very large compared to those which have been 
detected so far on the disk of Betelgeuse \citep[e.g.,][]{hpl09}.
Two PIONIER observations \citep{mkp15} are quasi-simultaneous with our Stokes $QU$ survey of \object{Betelgeuse}, 
namely on 11 January 2014 and 21 November 2014 (see Table\,\ref{tab1} for very near Narval's dates). 
The data concerning the three first PIONIER observations (January 2012, February 2013, January 2014) have been satisfactorily 
fitted with a model of disk with limb darkening (hereafter LDD model) and a Gaussian hot spot \citep{m14}. 
The spot is centered near the limb of Betelgeuse and its FWHM is about one radius of the star. 
The center of the spot has $\chi$ ranging from about 80$^\circ$ to about 110$^\circ$ which corresponds to the E-SE part of Betelgeuse's 
limb. 
For PIONIER data of 21 November 2014, the situation appears different and the fit of the model LDD $+$ Gaussian spot is worse 
than for the previous dates. 
In fact, a significantly better fit to this observation is obtained for a two-spot model. 
In this case the spot in the NW quadrant is the strongest one and has a Gaussian shape similar to that of the previous 
observations; the spot in the SE quadrant has a negligible diameter \citep{mkp15}.
Comparing the results from Table\,\ref{tab1} (with $\chi = \theta + 90^{\circ}$) and from the PIONIER observations, we see that the 
position of our spot1 in Set1 is similar to that of the large hot spot in January 2012, February 2013 and January 2014. For the November 2014 positions to be consistent, we have to choose for the polarimetric observation the location with $\chi$ 180$^\circ$ away the value reported in Table 1. Figure\,\ref{fig14} shows the PIONIER intensity maps presented by \cite{mkp15} and the maps obtained from polarimetry at the dates of quasi-simultaneous observations. This comparison suggests that the large hot spot observed by PIONIER  in 2012, 2013 and January 2014 is responsible for the stable polarization of our blue lobe from November 2013 to January 2014 and coincides with spot1. The evolution of the PIONIER hot spot observed in November 2014 could have the same origin as the evolution of spot1 observed from October 2014 to April 2015.

Our spot2 has not been detected with PIONIER. Our hypothesis that it is a spot located on the hidden hemisphere is one possible explanation. Also, \citet{d10}  proposed that the hot spots on \object{Betelgeuse} are composed of photospheric and chromospheric features. Dupree also proposed that these chromospheric features, are not detected by interferometry working in the infrared, as was the hot spot imaged by the HST/FOC \citep{gd96,u98}. 
Remarkably, the polarization angle of our redshifted lobe (linked to spot2) in 2015 is consistent with that (about 150$^\circ$) corresponding to the maximum of polarization observed by \citet{h84} and \citet{nrm94}, as reported by \citet{u98}.

\section{Conclusions}

We have discovered that most of the photospheric lines of \object{Betelgeuse} present a linearly polarized spectrum.
Together with the strong polarization of the \ion{Na}{i} D1 line, and including the polarization laws with both wavelength
and depth, we show that this linear polarization originates from the depolarization of the continuum.
Taking into account the discovery of linearly polarized spectra in the Mira star $\chi$\,Cyg and in the RV\,Tauri star R\,Sct 
\citep{laf15}, in which intrinsic polarization of resonance lines (similar to the Sun) is detected, all these observations 
of cool evolved stars point to a kind of second stellar spectrum. 

We interpret our results for \object{Betelgeuse} as being due to anisotropies of the radiation field induced by hot 
spots at the surface and Rayleigh scattering in the atmospheric envelope. 
We then propose an analytical model to interpret the observed polarization and we infer the presence of two hot spots 
and their positions on \object{Betelgeuse}. We show that applying the model to each velocity bin along the Stokes $Q$ and $U$ profiles allows us to derive a map of the involved bright spots on \object{Betelgeuse}. 
We compare our observations to quasi-simultaneous VLTI/PIONIER observations \citep{mkp15} and older measurements 
of the polarization of the continuum.
Our linear polarimetric observations suggest that we detected a hot spot seen by quasi-simultaneous PIONIER observations, 
and another one whose associated polarization angle implies a position near where the largest percentage of polarization of the continuum occurs  \citep[as reported by][]{u98}.

\begin{acknowledgements}
We thank the TBL team for providing service observing with Narval. We acknowledge the use of the database VALD (Vienna, Austria) and financial support from the "Programme National de Physique Stellaire" (PNPS) of CNRS/INSU, France. R.K.-A. acknowledges support by the Bulgarian-French mobility contract DRILA 01/3. GAW acknowledges Discovery Grant support from the Natural Sciences and Engineering Research Council (NSERC) of Canada.
\end{acknowledgements}

\end{document}